\documentclass[sigconf]{acmart}
\acmConference[CAIN 2024]{3rd International Conference on AI Engineering — Software Engineering for AI}{April 2024}{Lisbon, Portugal}%% Fonts used in the template cannot be substituted; margin 
\usepackage{xcolor}

\AtBeginDocument{%
  \providecommand\BibTeX{{%
    \normalfont B\kern-0.5em{\scshape i\kern-0.25em b}\kern-0.8em\TeX}}}
    
 \usepackage{enumitem}
\newcommand{\toolname}{VaryGen }
\begin{document}

\title{LLMs for Test Input Generation for Semantic Caches}

\author{Zafaryab Rasool, Scott Barnett, David Willie, Stefanus Kurniawan, Sherwin Balugo, Srikanth Thudumu, Mohamed Abdelrazek}
\email{{zafaryab.rasool, scott.barnett, david.willie, stefanus.kurniawan, s.balugo}@deakin.edu.au}
\email{{srikanth.thudumu, mohamed.abdelrazek}@deakin.edu.au}
\affiliation{%
  \institution{Applied Artificial Intelligence Institute, Deakin University}
  \city{Geelong}
  \country{Australia}}

% \email{firstname.lastname@deakin.edu.au}
% \thanks{\textsuperscript{1}Corresponding author: Zafaryab Rasool, email: zafaryab.rasool@deakin.edu.au.}

% \email{zafaryab.rasool, scott.barnett, david.willie, stefanus.kurniawan, s.balugo, srikanth.thudumu, mohamed.abdelrazek}@deakin.edu.au

\begin{abstract}
    
    Large language models (LLMs) enable state-of-the-art semantic capabilities to be added to software systems such as semantic search of unstructured documents and text generation. However, these models are computationally expensive. At scale, the cost of serving thousands of users increases massively affecting also user experience. To address this problem, semantic caches are used to check for answers to similar queries (that may have been phrased differently) without hitting the LLM service. Due to the nature of these semantic cache techniques that rely on query embeddings, there is a high chance of errors impacting user confidence in the system. Adopting semantic cache techniques usually requires testing the effectiveness of a semantic cache (accurate cache hits and misses) which requires a labelled test set of similar queries and responses which is often unavailable. In this paper, we present VaryGen, an approach for using LLMs for test input generation that produces similar questions from unstructured text documents. Our novel approach uses the reasoning capabilities of LLMs to 1) adapt queries to the domain, 2) synthesise subtle variations to queries, and 3) evaluate the synthesised test dataset. We evaluated our approach in the domain of a student question and answer system by qualitatively analysing 100 generated queries and result pairs, and conducting an empirical case study with an open source semantic cache. Our results show that query pairs satisfy human expectations of similarity and our generated data demonstrates failure cases of a semantic cache. Additionally, we also evaluate our approach on Qasper dataset. This work is an important first step into test input generation for semantic applications and presents considerations for practitioners when calibrating a semantic cache. 

\end{abstract}





\maketitle

% Using an LLM avoids the need to 1) annotate documents, 2) create a knowledge graph, or 3) return whole or partial documents in responses.Unlike traditional semantic search, LLMs support a conversational approach to search and generate human like responses. 

\section{Introduction}

Software development kits for creating applications that use large language models (LLMs) are starting to emerge i.e Microsoft's Semantic Kernel\footnote{https://learn.microsoft.com/en-us/semantic-kernel/overview/} and Langchain~\footnote{https://python.langchain.com/docs/get\_started/introduction}. These SDKs are growing in popularity as LLMs are able to 1) resolve ambiguities in natural language queries, 2) reason over unstructured documents, and 3) synthesise a human like responses to prompts \cite{bang2023gptcache}.  Semantic caches are becoming an important aspect of these SDKs as developers seek to create applications with a) lower latency, b) lower costs, and c) reduced environmental impact from using less compute resources (i.e. no repeat calls to LLMs with billions of parameters). 

A semantic cache enables similar queries (as defined by a similarity measure rather than an exact match) to match cached responses. However, testing the robustness of a semantic cache is challenging due to a) application specific calibration to balance the cache hits/misses ratio, b) domain specific query/response pairs, and c) a suitable similarity measure that handles the nuance in natural language (see \autoref{tab:incorrect-hit}). Ensuring the robustness of an application with a semantic cache relies on a representative sample of query/response pairs for calibration and testing.

\begin{table}[h]
    \centering
    \begin{tabular}{|p{2cm}|p{6cm}|}
    \hline
   \textbf{Query}& \textit{Do I have to reference in my Assignment 2?} \\
    \hline
   \textbf{Cached Query}& \textit{In Assignment 2, how many university references should I use?} \\
    \hline
    \end{tabular}
    \caption{An example of an incorrect cache hit using the semantic cache GPTCache\protect\footnotemark.}
    \label{tab:incorrect-hit}
    \vspace{-10pt}
\end{table}
\footnotetext{https://github.com/zilliztech/GPTCache}

% \begin{table}
%     \centering
%     \begin{tabular}{lp{5.5cm}}
%     \hline
%          \textbf{Query:}& Do I have to reference in my Assignment 2?\\
%          \textbf{Cached Query:}& In Assignment 2, how many university references should I use?\\
%          \hline
%     \end{tabular}
    
%     \caption[]{An example of an incorrect cache hit using the semantic cache GPTCache\footnotemark where a yes/no question matched with a quantity question.}
%     \label{tab:incorrect-hit}
%     \vspace{-7mm}
% \end{table}

% \footnotetext{https://github.com/zilliztech/GPTCache}

LLMs have been used to generate data for software applications \cite{li2023prompting, wang2023software, kang2023large, 10172490, alonso2022arte, mariani2014link} and for testing information retrieval based applications~\cite{alaofi2023can}. They base their work on the assumption of pre-existing questions which for new applications is not available. Recent work has also explored the use of LLMs for generating data for training Q\&A systems. However, their work focuses on the volume of questions from documents rather than subtle variations that would expose the limits of a semantic caching strategy. In summary, prior work has not considered the problem of test input generation for testing LLM based Q\&A systems. Our goal is to improve the robustness of Q\&A systems by test input generation and demonstrating the effectiveness of finding limitations in a semantic cache.

In this paper, we propose an innovative approach that uses LLMs for generating domain specific query/response pairs. We call this approach as Variation Generator (VaryGen). \toolname uses LLMs to a) generate queries, b) create hard negatives, and c) verify the quality of the generated pairs. We take a two stage approach, first we synthesise domain specific questions from a collection of documents, optionally including domain specific terms in the generation process. Second, from the synthesised questions we create variations resulting in hard negatives and close matches. To evaluate our approach, we 1) conducted a qualitative analysis of the synthesised questions (N=100), and 2) through a case study of an open source semantic cache, GPTCache. Our results found limitations in the default configurations of GPTCache demonstrating that our approach is suitable for test input generation. However, the qualitative evaluation revealed additional filters are required to improve the variations of questions which we leave for future work.

% However, the qualitative evaluationhighlighted the necessity for additional filters to enhance question diversity, left for future exploration

% However, the qualitative evaluation revealed additional filters are required to improve the variations of questions which we leave for future work. 

The contributions arising from this work are:
\begin{itemize}
    \item An automated approach for synthesising query/response pairs that contain hard negatives. This approach has been designed for testing applications that include QandA or a semantic cache for natural language queries. 
    \item A qualitative analysis of the challenges with using LLMs for synthesising test input data. 
    \item An empirical case study with an open source semantic cache, GPTCache. This showed the nuances in similarity measures that software engineers need to test to improve the robustness of their applications. 
\end{itemize}

\begin{figure*}[h]
    \centering
    \includegraphics[width=1\linewidth, height=6.5cm]{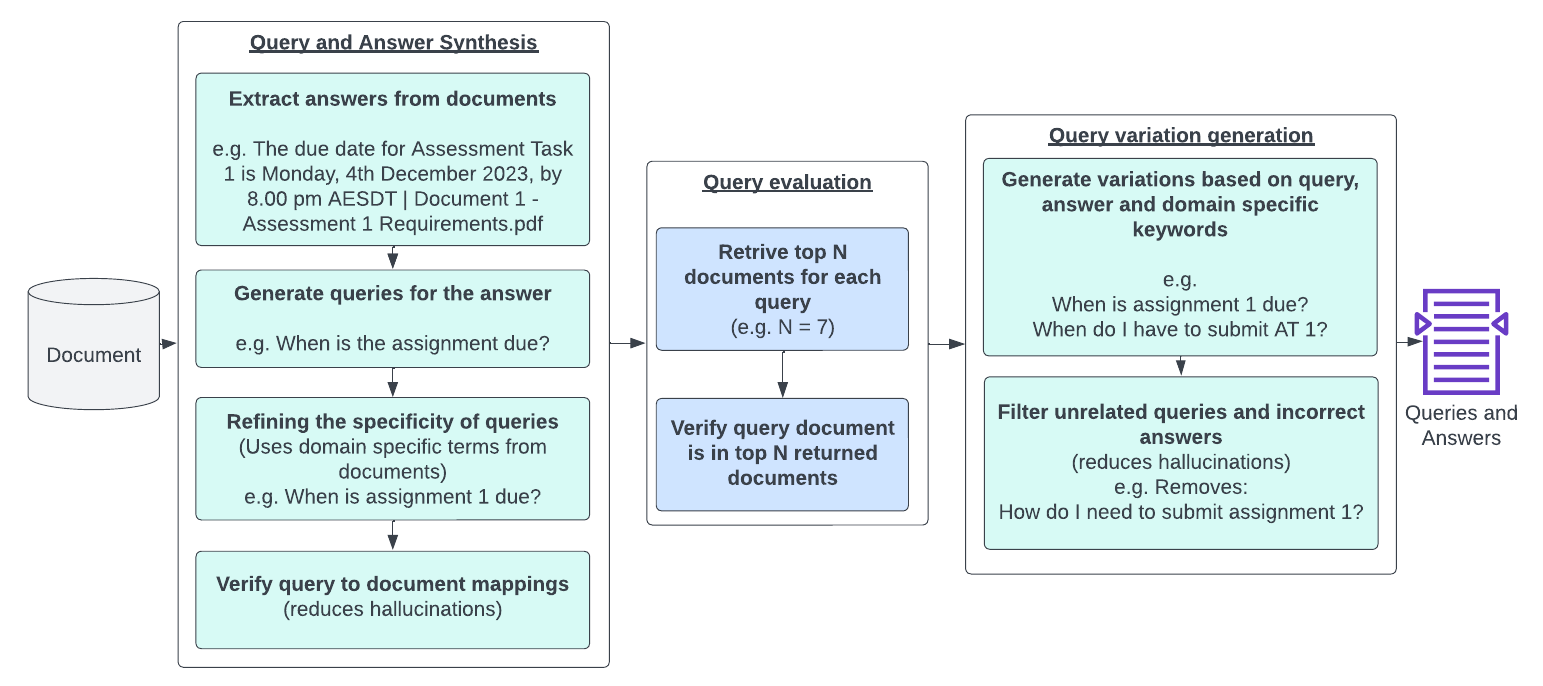}
    \caption{The 3 stages in \toolname : 1) Query Generation, 2) Query Evaluation and 3) Query Variation Generation. Green boxes indicate steps that involve a LLM and blue boxes that are automated without the use of an LLM. }
    \vspace{-5pt}
    \label{fig:approach-diagram}
\end{figure*}

\section{Semantic Cache}
\label{sec:motivation}
For the purpose of this paper, we investigate the use of a semantic cache as part of a question and answering system where the queries are questions and the answers are the responses. Semantic caching has been a long studied problem \cite{godfrey1999answering,ahmad2020enhanced,ajarroud2022sbqp, salas2022semantics, frieder2022caching} for efficient retrieval of query responses by reusing already extracted data. A semantic cache is designed to optimise the performance and efficiency of systems that process large volumes of data, such as those based on LLMs. Unlike traditional caches that store data based on the exact queries or data requests, a semantic cache understands the meaning or "semantics" of the data and the queries. This allows it to not just store exact query results, but also data that is semantically related to these queries. The semantics are represented by a text embedding, an array of floating point values that has been trained with a language model. When a new query is made, the semantic cache retrieves data that is contextually relevant, even if the new query is not an exact match to previous ones. 

Text embeddings are imprecise for subtle variations in natural language and for domain specific terminology. When using a semantic cache, a search algorithm such as K-nearest neighbours is used to find relevant embeddings before filtering to a single result. This approach requires calibration as it will always return a result even if the answers aren't similar. The robustness of an LLM based application requires a well calibrated semantic cache. Incorrect configuration results in incorrect cache hits (i.e. the user receives an incorrect answer), or many incorrect cache misses - mitigating the benefits of using a semantic cache. Each application requires calibration due to domain specific terminology and diversity of content i.e. the quality of the embeddings for each domain is unknown until tested. To calibrate a semantic cache a representative dataset is required that includes both true matches and almost but incorrect matches.

Thus a test input generation approach for semantic caches requires the following:

\begin{itemize}
    \item Generate domain relevant queries automatically. 
    \item Include a test oracle to ensure that generated queries are relevant to the domain. 
    \item Create semantically "close" variations of queries (hard negatives and true pairs). 
    % \item Create diverse samples that cover all of the domain. 
    \item Automated to scale the data generating process to support accurate calibration. 
\end{itemize}

\section{\toolname}

% \todo{Make sure that this section describes how the requirements at the end of \section{sec:motivation} are satisfied by our approach. Ideally, the evaluation should cover some or all of the requirements too.}

% \todo{Make sure that this section describes how the requirements at the end of Motivation section are satisfied by our approach. Ideally, the evaluation should cover some or all of the requirements too.}

Based on the discussion above, we propose a test input generation approach using LLM. This 3-step approach, (detailed below) depicted in Figure \ref{fig:approach-diagram}, combines the strengths of LLMs, domain knowledge, and rigorous evaluation to generate diverse, contextually relevant, and high-quality test input data for semantic applications. 

% In this section, we explain our approach for query generation and evaluation. This 3-step approach (detailed below) combines the strengths of LLMs, domain knowledge, and rigorous evaluation techniques to generate diverse, contextually relevant, and high-quality test input data for semantic applications. 

% \subsection{Motivation}

\subsection{Query And Answer Synthesis}

In formulating our approach, we made three assumptions: 1) a corpus of documents is available, 2) the documents are a semantic representation (i.e. an embedding vector that represents the semantic content of the document), and 3) a way to retrieve answers to queries from the documents is available. 

The initial step involves extracting relevant answers from the provided document. The process of answer extraction from documents for test input generation involves utilizing LLMs and a document corpus and is as follows: 

First, we provide a document to an LLM and prompt it to extract facts from the document. Most documents consist of multiple facts or sentences that can serve as an answer to specific questions. Thus, we use these facts as answers for generating questions. Note that, if the document is too long or consists of multiple documents, a retriever can be used to extract relevant facts from it. However, for simplicity, we provide full document passages in the prompt to an LLM to extract answers. 
We then again prompt the LLM to generate questions for each of the extracted answer.  Here, proper prompting and using extracted answers restrict the generated questions to the scope of the document only, thereby obtaining domain relevant queries automatically.

\subsection{Query Evaluation}
The generated queries obtained in the previous step need to be verified for correctness and whether they align contextually with the content of the document, which is the aim of this step.

This step validates the accuracy of the generated queries by mapping them back to the original document. For each query, the top $N$ documents are retrieved and assessed whether the original document is included in the returned set. The queries that extract relevant documents are included while the others are excluded from the set. The refined set of generated queries is then used in the next step to generate variations of the queries.

Thus, this verification step ensures that the obtained queries align contextually with the content of the document (i.e., they are domain specific), establishing a reliable connection between the generated questions and the source material.

% This step retrieves the top $N$ documents for each query and assess whether the original document is included in the returned set. The queries that do not fit the criteria i.e. that do not extract the relevant documents are excluded from the set. 

% This step gauges the relevance and informativeness of the queries in retrieving pertinent information. 

\begin{table}[]
    \centering
    \begin{tabular}{p{8cm}}
    \hline
    Question: \{question\}\\
    Answer:  \{answer\}\\
    \\
     Generate variations ONLY from the provided Question above and ONLY use the Answer to constrain the generated questions.  \\
     \\
     Variation Guidelines: \{....\}\\
     Response:
     \\
     \hline
    \end{tabular}
    \caption{Prompt template for Query Variation Generation}
    \label{tab:prompt_varygen}
    \vspace{-20pt}
\end{table}

\subsection{Query Variation Generation}
This step focuses on generating semantically similar variations of the queries obtained in the previous step. The idea is to generate as diverse variations of the query as possible while still maintaining semantic similarity. 

% For generating queries that are diverse while still semantically similar, variations of the generated queries need to be produced. 

We prompt the LLM to generate variations of the queries making sure that: (i) the queries are diverse i.e. different in terms of phrasing, structure, or specific keywords, and (ii) any unrelated or hallucinated queries are removed. This maintains the quality of the generated variations. We provide both the question and the answer in the prompt in order to 
keep the queries related to the extracted content and that the domain knowledge is reflected in the generated variations. 

Utilising proper prompting, this step refines the queries by incorporating specific terms, concepts, or structures relevant to the domain, ensuring that the generated test inputs align closely with the intricacies of the subject matter. Table \ref{tab:prompt_varygen} provides an example of a prompt used to achieve the above objective. The guidelines help to provide relevance and specificity in the query variations. 

Thus, using the above approach, we get the required query variations that can be used to test semantic applications. Next, we discuss the evaluation of our input data generation approach.

% The diverse queries should include queries that are different in terms of phrasing, structure, or specific keywords, while any unrelated or hallucinated queries be removed to maintain the quality of the generated variations.

% Using proper prompting, we utilise LLM's natural language understanding to create diverse queries related to the extracted content. Furthermore, to enhance the specificity and relevance of the generated queries, domain knowledge is employed. This step refines the queries by incorporating specific terms, concepts, or structures relevant to the domain, ensuring that the generated test inputs align closely with the intricacies of the subject matter. Table \ref{tab:prompt_varygen} provides the example of prompt used to achieve the above objective.  Using given $question$ and $answer$, an LLM is asked to generate variations of query using the guidelines provided. The guidelines, as stated earlier, help to provide relevance and specificity.

% \subsection{Variation Evaluation}

% The final step involves evaluating the query variations. 

% The LLM is prompted to answer the generated variations, and open AI evaluations are applied to a sample of questions. This evaluation process ensures that the variations maintain coherence with the original queries and adhere to the expected standards of correctness, relevance, and domain-specific nuances.

\section{Preliminary Evaluation}

In this section, we discuss the dataset details and the evaluation strategy along with the results of the evaluations. 

\subsection{Dataset Description}
We use two datasets for our study. 
The Qasper dataset \cite{dasigi2021dataset} is an information-seeking dataset consisting of research papers on natural language processing. The dataset contains question-answers with different question types such as abstractive, extractive, boolean and unanswerable.  We filter the dataset by removing the unanswerable ones for our purpose.

The Assignment dataset is a manually curated dataset containing 61 questions crafted from sample questions representing what students are likely to ask during a university subject. The dataset consists of questions regarding the assignments such as word limit, requirements, and use of GenAI. As some questions can be similar, we further excluded questions that are too similar by manual inspection, leaving only 52 unique questions.

\subsection{Qualitative Evaluation}

% \begin{table}[]
%     \centering
%     \begin{tabular}{|c|c|c|c|}
%     \hline
%     \textbf{Dataset} & \textbf{Topic} & \textbf{Correct} & \textbf{Incorrect} \\
%     \hline
%     Assignment & Unit-related Questions     & 80 & 20 \\
%     \hline
%     Qasper & NLP   & 84 & 16 \\
%     \hline
%     \end{tabular}
%     \caption{Qualitative evaluation of the query variations.}
%     \label{tab:results}
% \end{table}

The focus of \toolname was to create an automated query and answer generation pipeline to support software engineers working with a semantic cache. We conducted evaluations in two phases to ensure a thorough analysis of the \toolname approach. These evaluations covering the different steps of \toolname ensure representative samples of each of the core stages is evaluated.

% The first evaluation comprise human evaluation of the generated questions. We randomly sampled 100 questions and checked if the answers seem relevant to the generated questions. If the answer was relevant to the question and if the question was realistic in the sense that students would ask such questions, we marked it as True, otherwise False. 
% The second evaluation comprise human evaluation of the query variations generated for each query. Again, we randomly sampled 100 question-answer pairs and checked each question for correctness by comparing it with the original answer. In other words, we check if the answers made sense for those questions. 

\textbf{Method:} The initial evaluation involved a human assessment of the generated questions. We conducted a random sampling of 100 questions, scrutinizing each to determine if the associated answers appeared relevant to the generated questions. Our criterion for marking a question as \lq Correct' included assessing both the relevance of the answer to the question and the overall realism of the question – whether it resembled the type of questions students would typically ask. Questions meeting these criteria were marked as \lq Correct', while those that did not were labelled as \lq Incorrect'. We conducted an initial evaluation for the Assignment dataset as it consists of documents and not for the Qasper dataset, which consisted of questions that could be directly used for the next step.

The second phase of evaluation involved a human assessment of the query variations generated for each query. In this stage, we randomly selected 100 question-answer pairs and examined each question's correctness by comparing it against the original answer. Essentially, our focus was on verifying whether the generated questions aligned sensibly with the corresponding answer. This evaluation is done for both datasets.

\textbf{Results:} 
The results of the initial evaluation on the Assignment dataset indicate that 81\% of the generated questions were realistic, while only 19\% did not satisfy our criteria. This demonstrates that our approach generates relevant questions that are domain-specific and realistic. For example, questions such as "What do I need to formulate that is original and related to the topic I have chosen for Assessment Task 2?" may not be asked by a student. We now discuss the evaluation results on query variations.  For the Assignment dataset, we found that 80\% of the sampled query variations were correct, while 20\% were incorrect. Similarly, for the Qasper dataset, the correct was 84\%  while the incorrect \% was only 16. 

% This demonstrates that most query variations generated using our approach are correct.

These results from the evaluations emphasise that most generated questions and the variations of the query for both datasets were semantically similar to the original questions from which the variations were generated. 

\begin{table}[!t]

    \centering
    \begin{tabular}{p{1.5cm}p{1.2cm}p{1.2cm}p{1.2cm}p{1.2cm}}
    \hline
           \textbf{Strategy} 
           &  \textbf{Correct} \textbf{Hits} 
           &  \textbf{Incorrect}  \textbf{Hits}
           &  \textbf{Correct} \textbf{Misses}
           &  \textbf{Incorrect} \textbf{Misses}
           \\
            \hline
           DistilBERT & 1234 & 885 & 19 & 128 \\
           SBERT & 1422 & 360 & 38 & 446 \\
           ONNX & 1111 & 314 & 35 & 806 \\
        \hline
    \end{tabular}
    \caption{Results from evaluating the generated dataset against three semantic caching strategies}
    \label{tab:gptcache_results}
    \vspace{-20pt}
\end{table}

\subsection{Semantic Cache Case Study}
We analysed the performance of an open source semantic cache library using the generated dataset.
% -- Dave, please add details and verify

\textbf{Method.} For the purpose of our case study, we evaluated GPTCache \cite{bang2023gptcache}, an open source semantic cache library with various semantic caching strategies. GPTCache was selected due to a) intended use for LLM applications, and b) popular (over 5000 stars). We used the Assignment dataset and evaluated the performance of GPTCache using three strategies with a similarity threshold of 0.9 (selected through empirical evaluation):

%[topsep=0pt, itemsep=-1pt,leftmargin=*]
\begin{itemize}
    \item DistilBERT (base, uncased) embeddings with Cosine Similarity for similarity evaluation
    \item SBERT embeddings with GPTCache's SBERT CrossEncoder implementation for similarity evaluation
    \item ONNX embeddings with GPTCache's ALBERT ONNX implementation for similarity evaluation
\end{itemize}

The following metrics were used to measure the performance of GPTCache: Number of correct cache hits, Number of incorrect cache hits, Number of correct cache misses and Number of incorrect cache misses. Correct Hits indicate a correct value was retrieved from the cache. Incorrect Hits indicate a non-matching value was returned. Correct Misses indicate no value was returned when no value was expected, and Incorrect Misses indicate a value was not returned where one was expected.

\textbf{Results.} We used the Assignment dataset consisting of a total of 2266 questions including the original questions and their variations generated using our approach. We put the dataset of questions into the cache, one by one, and measure the above metrics. The results of our evaluation are shown in Table \ref{tab:gptcache_results} which demonstrates the generated question set was capable of providing examples that resulted in Correct hits, Incorrect Hits and Incorrect Misses on all three caching strategies.

% True Positive - Correct Cache Hit
% \begin{table}[!t]
%     \centering
%     \begin{tabular}{p{1.2cm}p{3.2cm}p{3.2cm}}
%     \hline
%     % \multicolumn{3}{c}{\textbf{True Positive - Correct Cache Hit}} \\ 
%     \textbf{Strategy} & \textbf{Query} & \textbf{Cached Key} \\
%     \hline
%     DistilBERT & What is the word limit for each essay in AT3 without the 10\% buffer? & What's the word limit for each essay in AT3 with no 10\% buffer? \\
%     SBERT & What is the list of injunctions to be followed for Assignment 1? & What is the list of injunctions and prohibitions to be followed for Assignment 1? \\
%     ONNX & How many words should I use when writing about my use of generative AI for Assignment 1? & How many words should I use explaining my use of generative AI for Assignment 1? \\
%     \hline
%     \end{tabular}
%     \caption{True Positive - Correct Cache Hit}
%     \label{tab:true_positive}
% \end{table}

\begin{table}[!t]
    \centering
    \begin{tabular}{p{1.2cm}p{3.2cm}p{3.2cm}}
    \hline
    % \multicolumn{3}{c}{\textbf{True Positive - Correct Cache Hit}} \\ 
    \textbf{Strategy} & \textbf{Query} & \textbf{Cached Key} \\
    \hline
    DistilBERT & For the success of this unit, is the use of generative AI necessary? & Does this unit require to use generative AI? \\
    SBERT & In Assignment 2, how many academic citations are necessary? & How many scholarly references should be included in Assessment 2? \\
    ONNX & How extensive should my description of using generative AI be for Assessment 1? & How comprehensive should I be when explaining my use of generative AI for Assignment 1? \\
    \hline
    \end{tabular}
    \caption{True Positive - Correct Cache Hit}
    \label{tab:true_positive}
    \vspace{-15pt}
\end{table}

% False Positive - Incorrect Cache Hit
\begin{table}[!t]
    \centering
    \begin{tabular}{p{1.2cm}p{3.2cm}p{3.2cm}}
    \hline
    % \multicolumn{3}{c}{\textbf{False Positive - Incorrect Cache Hit}} \\ 
    \textbf{Strategy} & \textbf{Query} & \textbf{Cached Key} \\
    \hline
    DistilBERT & What is the word limit for the background section of my Assessment 1? & What is the recommended word count for the aims section of my report for AT1? \\
    SBERT & What is the directive for Assignment 1? & How many words should go into my campaign strategy for Assignment 1? \\
    ONNX & How should I proceed with Assessment 2? & How can I get a high distinction in Assessment 2? \\
    \hline
    \end{tabular}
    \caption{False Positive - Incorrect Cache Hit}
    \label{tab:false_positive}
    \vspace{-15pt}
\end{table}

% True Negatives - Correct Cache Miss
% \begin{table}[!t]
%     \centering
%     \begin{tabular}{p{1.2cm}p{6.6cm}}
%     \hline
%     % \multicolumn{2}{c}{\textbf{True Negatives - Correct Cache Miss}} \\
%     \textbf{Strategy} & \textbf{Query} \\
%     \hline
%     DistilBERT & Does this unit require to use generative AI? \\
%     SBERT & What do I have to do to get a high distinction in Assignment 3? \\
%     ONNX & How much is Assignment 1 worth? \\
%     \hline
%     \end{tabular}
%     \caption{True Negatives - Correct Cache Miss}
%     \label{tab:true_negatives}
% \end{table}

% False Negatives - Incorrect Cache Miss
\begin{table}[!t]
    \centering
    \begin{tabular}{p{1.2cm}p{3.2cm}p{3.2cm}}
    \hline
    % \multicolumn{3}{c}{\textbf{False Negatives - Incorrect Cache Miss}} \\ 
    \textbf{Strategy} & \textbf{Query} & \textbf{Expected Key} \\
    \hline
    DistilBERT & What should the layout of my AT1 look like? & What sections should I have in my AT1? \\
    SBERT & What's the brief for Assignment 1? & What do I need to do for Assignment 1? \\
    ONNX & What methodologies are feasible for Assignment 2? & What methods can I use for Assignment 2? \\
    \hline
    \end{tabular}
    \caption{False Negatives - Incorrect Cache Miss}
    \label{tab:false_negatives}
    \vspace{-20pt}
\end{table}

% in 
% Tables \ref{tab:true_positive}, \ref{tab:false_positive}, \ref{tab:false_negatives} along with the caching strategies used. 

We discuss some sample examples of the above results.
% Correct cache hits indicate that our generated queries align well with the stored information, validating the relevance and accuracy of our synthesized questions, 
For correct cache hits, as shown in Table \ref{tab:true_positive}, all three sample query, while different in structure and phrasing, are similar in meaning to the respective cached key, which validates the relevance and accuracy of the synthesized questions. This highlights the robustness and effectiveness of our test input generation approach in capturing semantic relevance across many different formulations of queries.

% Regarding incorrect cache hit examples shown in Table \ref{tab:false_positive}, in the first example, the generated query is seeking information about the word limit for the background section and the cached query is related to the word count for the aims section for Assessment 1.

Regarding incorrect cache hit examples shown in Table \ref{tab:false_positive},
the first example shows that although both queries pertain to Assessment 1, they focus on different aspects (background vs. aims), making them semantically distinct. The reason for this incorrect cache hit could be attributed to the limitations in the semantic cache's ability to discern subtle variations in the meaning of queries. The second and third examples are also semantically different in terms of their requirements and follow similar explanations. 
% This emphasizes the importance of refining the specificity and context awareness of queries to enhance the performance of semantic caching systems. 

% While the generated query is contextually different, the semantic cache may not possess the fine-grained understanding to accurately distinguish between queries that are similar but refer to different sections of the assessment. 

Regarding incorrect cache miss examples shown in Table \ref{tab:false_negatives}
in the first example, the semantic cache fails to identify the correct match, as it may not fully capture the specific intent (sections and layouts) behind the generated query. For the second example, the generated query is seeking information about the brief or instructions for Assignment 1, while the expected cached query is a more general inquiry about the tasks required for Assignment 1. The semantic cache fails possibly due to the subtle differences in the queries semantics. Similarly, for the third example, the terms "methodologies" and "methods" may not have a direct one-to-one correspondence in the semantic embeddings.

% The above results shows that our approach is capable of uncovering scenarios where the semantic cache strategies may fail to identify relevant matches or may incorrectly retrieve matches. This demonstrates the utility of our approach in revealing potential challenges and areas for refinement in the specificity, diversity, and coverage of the generated queries to enhance the performance of semantic caching systems. Additionally, these results also provide valuable insights for refining semantic caching strategies.

Thus, our approach is capable of uncovering scenarios where the semantic cache strategies may fail to identify relevant matches or may incorrectly retrieve matches. This demonstrates areas for refinement in the specificity, diversity, and coverage of the generated queries to enhance the performance of semantic caching systems. 

\section{Related Work}

Recent studies have demonstrated the potential for leveraging LLMs to automatically generate various test artifacts including test cases, test inputs, and test oracles \cite{wang2023software}. Kang et al. \cite{kang2023large} proposed an approach for using an LLM to reproduce bugs and generate reliable test cases to suggest to developers. Yu et al. \cite{yu2023llm} demonstrated how well LLMs can adapt to diverse systems while creating test scripts. Liu et al. \cite{10172490} introduced an LLM-based approach to intelligently generate semantic input text based on GUI context. Other works have focused on using LLMs or semantic knowledge bases to generate valid test inputs. Alonso et al. \cite{alonso2022arte} proposed ARTE, which uses knowledge bases such as DBPedia to extract realistic web API test inputs based on API specifications. Mariani et al. \cite{mariani2014link} introduced Link, which analyses a GUI's input fields, queries DBPedia to find matching data, and generates complex test inputs. Wanwarang et al. \cite{wanwarang2020testing} leveraged both static and dynamic analysis along with an LLM to generate test inputs. Some studies have investigated using LLMs to generate test oracles, which determine the correctness of test outputs. Yang et al. \cite{yang2023beyond} used an LLM to create knowledge bases to guide testers on what outputs to validate. 
Alaofi et al. \cite{alaofi2023can} showed LLMs can generate variations of queries to create test oracles.
% create test oracles. 
% LLMs have supported other aspects of testing as well. 
% Li et al. \cite{li2022self} used an LLM for hard negative generation to improve person re-identification.

LLMs show promise for the automated generation of various test artefacts, reducing manual effort while improving test coverage and effectiveness. However, there remain open challenges in improving the reliability and semantic correctness of LLM-generated tests.

\section{Limitations and Future Research Directions}

While our approach demonstrates promise in generating domain-specific query/response pairs for testing LLMs and semantic applications, certain limitations underscore areas for future research. The need for additional filters to enhance the quality of query variations, identified through our evaluation, prompts further investigation into advanced techniques for refinement. Additionally, our approach currently lacks human validation during data generation, which means some generated questions may be of lower quality or incorrect. However, we argue that an automated, inexpensive dataset that uncovers software limitations, as shown in our case study, still provides utility. Incorporating human annotation to improve dataset quality remains an area for future work. 

% Additionally, the dependency on manual calibration for semantic caches highlights the potential for automated methods to adapt to diverse domains more seamlessly. 

Additionally, the inherent challenge of subtle semantic nuances and imprecisions in text embeddings requires attention, urging the exploration of methods to fine-tune LLMs or incorporate domain-specific embeddings. 
Our evaluation also focused on a single semantic cache, so further empirical comparisons of multiple semantic caching systems should be carried out, which is now feasible using \toolname. Other potential areas for future work include performing ablation studies to reduce data generation costs for increased scale, defining quality models and metrics to improve query generation, additional empirical evaluations of open source caches, generating datasets to produce guards protecting semantic applications, incorporating diversity metrics to broaden the scope of questions, and further analysis of how synthesis techniques could be tailored and improved for diverse domains and applications.

Addressing the limitations and exploring these research directions will contribute to the continual improvement and advancement of test input generation methodologies for LLMs and semantic applications.

\section{Conclusion}
This paper has explored the evolving role of software development kits (SDKs) that leverage large language models (LLMs) to enable more advanced natural language applications. Semantic caches have emerged as a pivotal component within these SDKs, aiming to reduce latency, costs, and environmental impact by avoiding redundant calls to the LLM. However, the imprecision of text embeddings requires careful calibration of semantic caches to prevent incorrect cache hits. To address this, we have proposed and evaluated an innovative approach for automated test input generation using LLMs themselves. Through a two-stage process of synthesising domain-specific queries and creating semantically close variations, including hard negatives, representative test data can be generated at scale. Both the qualitative analysis and semantic cache case study validate the suitability of this approach for the robust testing necessary to ensure cache calibration. This research contributes a holistic perspective, not just on test input generation, but on the end-to-end robustness and calibration needs of semantic caches in this evolving landscape of LLM-based SDKs. The proposed approach paves the way for enhanced software development practices that fully harness advanced language models while safeguarding reliability.

\section{Acknowledgements} 
To Rajesh Vasa and Kon Mouzakis for their insightful guidance and unwavering support that greatly enriched this research project.
\bibliographystyle{ACM-Reference-Format}
\bibliography{references}

\appendix


\end{document}